\documentclass[aps,epsf,preprint,twocolumn]{revtex4}
\usepackage{latexsym}
\usepackage{amsfonts}
\usepackage{graphicx} 
\usepackage{epsfig}

\begin{document}

\begin{center}
{\bf \large \large Breakdown of Stokes-Einstein relation in
supercooled water}
\end{center}

\medskip

\begin{center}
{\bf Pradeep Kumar}
\bigskip

\noindent{Center for Polymer Studies and Department of Physics, Boston
University,~Boston, MA 02215 USA}
\end{center}
\hfill Aug 9, 2006
\medskip


Almost everyone would agree that water is probably the most important
liquid for life, but few fully appreciate that water is also the most
puzzling among the liquids.Water is anomalous in many ways compared to
simple liquids. One of the most well known anomalies is the decrease
of density of water upon cooling below $4^{o}$~C. Other anomalies of
water include the increase of specific heat and compressibility upon
cooling. 

Apart from the thermodynamic anomalies, the dynamics of water also
displays surprising properties, such as the increase of diffusivity
and decrease of viscosity upon compression~\cite{debenedetti} and a
breakdown of the Stokes-Einstein (SE) relation~\cite{einstein} in
supercooled water. The SE relation, is a hydrodynamic equation
relating the diffusivity $D$, the temperature $T$, and relaxation time
$\tau$ (assuming that $\tau$ is proportional to the
viscosity). Specifically, it states that $D$ is proportional to $T$
and inversely proportional to the relaxation time $\tau$.  Hence the
product $D\tau/T$ should be temperature independent.  This important
hydrodynamic relation is obeyed by many liquids~\cite{OTP} at
relatively high temperatures, but usually fails to describe the
dynamics in supercooled states. Breakdown of the SE relation has also
been seen in simple liquids near the glass transition
~\cite{gotze,OTP,berthier}. It was hypothesized that the presence of
large dynamic heterogeneities might cause the breakdown of SE
relation~\cite{ediger,ngai}. Dynamic heterogeneities in liquids are
the consequence of highly mobile molecules forming a cluster and
moving coperatively. The size of these spatially heterogeneous
dynamical regions increases as the temperature is decreased. Dynamic
heterogeneities facilitate diffusion ( and local strucutral relaxation
), but do not facilitate the relaxation of the entire system. Hence
large size of these dynamic heterogeneities should ``decouple''
diffusion and relaxation so $D\tau/T$ is no longer a constant. The
presence of dynamic heterogeneities in supercooled liquids has been
verified by both experiments~\cite{berthier,weeks} and computer
simulations~\cite{glotzer}. In molecular liquids as well as in
colloidal suspensions, the glass transition is followed by a sharp
growth in dynamic heterogeneities~\cite{berthier,weeks}. Hence it is
common to assume that SE breakdown is related to glass transition
since the sizes of dynamic heterogeneities grows sharply near the
glass transition~\cite{chandler,OTP,tarjus,ngai}.

The work by Chen et. al. in this issue of PNAS~\cite{chen-SE} on water
confined in nanopores, sheds light on the long standing issue of
decoupling of diffusion and structural relaxation in supercooled
water. They measure the diffusivity $D$ by NMR experiments and the
relaxation time $\tau$ by neutron scattering experiments for
temperatures down to $-83^{o}$~C or 190~K. Confinement prevents the
freezing of water below the bulk homogeneous nucleation temperature of
$-38^{o}$~C. Chen et. al. find that the product $D\tau/T$ is a
constant at higher temperatures, but increases sharply at low
temperatures, signaling the breakdown of the SE relation. They further
find that the breakdown of the SE relation occurs well above water's
glass transition temperature~\cite{chen-SE}. Hence their experiments
call out for a scenario different from that believed for simple
colloidal suspensions and liquids~\ref{OTP}. A possible clue for this
new scenario arises from the fact that Chen et. al. ~\cite{chen-SE}
discover that the SE relation breaks down at the same temperature
where the behavior of the dynamics of water changes from non-Arrhenius
at high temperatures to Arrhenius at low temperatures.



\begin{figure}[t]
\begin{center}
\includegraphics[width=8cm]{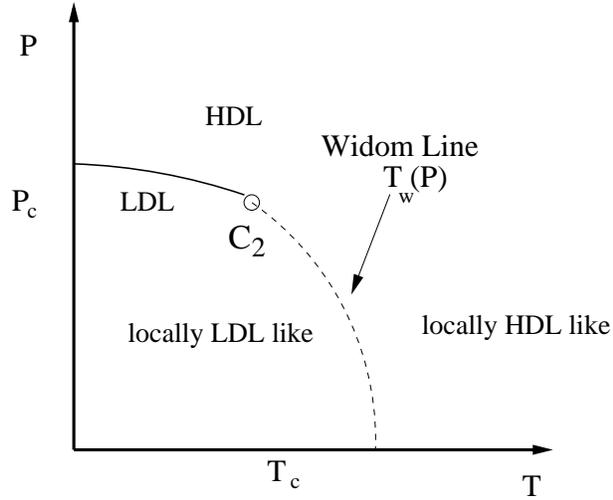}
\end{center}
\caption{Low temperature liquid-liquid critical point $C_2$ of water
and corresponding Widom line. The coexistence line between LDL and HDL
is represented by a solid curve while the Widom line $T_W(P)$
emanating from the LL-critical point is represented by a dotted
curve. Widom line separates the water with HDL-like feature at high
temperature from the LDL-like feature at low temperatures.}
\label{fig:widom}
\end{figure}

\begin{figure}[t]
\begin{center}
\includegraphics[width=6.5cm]{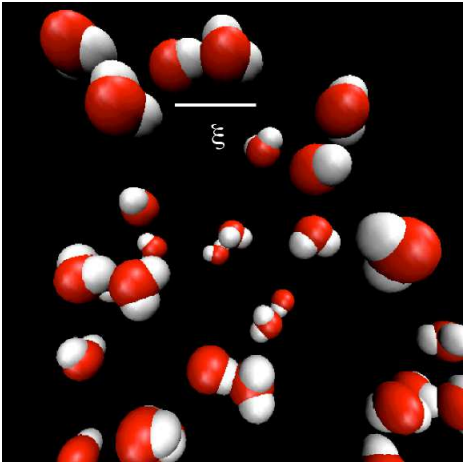}
\includegraphics[width=6.5cm]{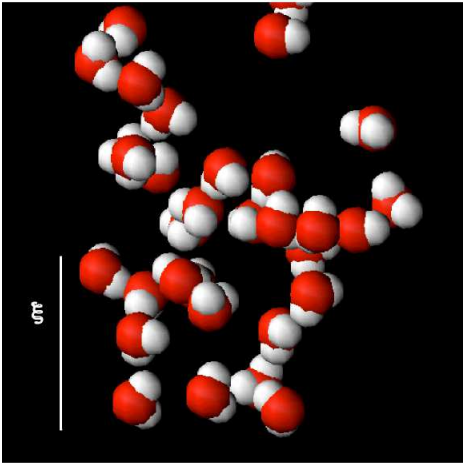}
\end{center}
\caption{ Effect on the size of dynamic heterogeneities on crossing
  the Widom line $T_W(P)$. Clusters of $7\%$~ most mobile molecules
  showing the dynamic heterogeneities (a) for $T>T_W(P)$ and (b) for
  $T < T_W(P)$. The characteristic size of the dynamic heterogeneities
  $\xi$ increases sharply upon crossing $T_W(P)$~ from high
  temperature side to low temperature side because of the emergence of
  a locally LDL-like structure.}
\label{fig:DH-water}
\end{figure}

One possible interpretation of the breakdown of SE relation in
supercooled water consistent with their experimental
results~\cite{Liu04} is the presence of a hypothesized liquid-liquid
critical point~\cite{mishima98,poole1} in the deep supercooled region
of water. The liquid-liquid critical point gives rise to a Widom line
$T_W(P)$ (see Fig.~\ref{fig:widom}), which is the locus of the
correlation length maxima, which emanates from the LL-critical point
and extends into the one-phase
region~\cite{AnisimovbookB,AnisimovbookC,xuPNAS}. When water is cooled
along a constant pressure path below the critical point, it changes
from a predominance of a local HDL structure to a predominance of a
local LDL structure upon crossing the Widom line. Two consequences of
this interpretation are:

\begin{itemize}

\item {The dynamics in the HDL-like region above $T_W(P)$  is
  expected to be non-Arrhenius while the dynamics in LDL-like region
  is expected to be Arrhenius. These expectation are borne out by the
  experiments~\cite{Liu04,xuPNAS,mallamaceJCP}}.

\item {In the HDL-like region diffusion is a cooperative phenomenon
    but with small length and time scales of dynamic heterogeneities,
    so we expect the SE relation to hold, while in the LDL-like
    region, emergence of a large ``ice-like'' local structure below
    $T_W(P)$ causes a growth of dynamic heterogeneities. Hence in
    the LDL-like region the diffusion would be more cooperative
    compared to the HDL-like region. Indeed we find that the size of
    the dynamic heterogeneities has a sharp increase at
    $T_W(P)$. Fig.~\ref{fig:DH-water} shows the formation of clusters
    by most mobile molecules of water in computer simulations above
    (Fig.~\ref{fig:DH-water}(top panel)) and below
    (Fig.~\ref{fig:DH-water}(bottom panel)) $T_W(P)$. The size $\xi$
    of these dynamic heterogeneities increases as the temperature is
    lowered and has a sharp increase at the Widom line temperature
    $T_W(P)$ and thus a breakdown of the SE relation.}

\end{itemize}

In summary, the discovery of Chen et.~al.~\cite{chen-SE} that the
    breakdown of the SE relation in supercooled water does not occur
    near the glass transition temperature raises the question of what
    causes the SE breakdown in supercooled water. It is possible that
    the cause is crossing the Widom line which arises from the
    liquid-liquid critical point.

We thank the NSF for support.

%


\end{document}